\documentclass[preprint,showpacs,preprintnumbers,amsmath,amssymb]{revtex4}
\usepackage{epsfig}
\usepackage{graphicx}
\usepackage{dcolumn}
\usepackage{bm}
\usepackage{amsmath}
\usepackage{subfigure}

\let\jnfont=\rm
\def\NPB#1,{{\jnfont Nucl.\ Phys.\ B }{\bf #1},}
\def\PLB#1,{{\jnfont Phys.\ Lett.\ B }{\bf #1},}
\def\EPJC#1,{{\jnfont Eur.\ Phys.\ Jour.\ C }{\bf #1},}
\def\PRD#1,{{\jnfont Phys.\ Rev.\ D }{\bf #1},}
\def\PRL#1,{{\jnfont Phys.\ Rev.\ Lett.\ }{\bf #1},}
\def\MPLA#1,{{\jnfont Mod.\ Phys.\ Lett.\ A }{\bf #1},}
\def\JPG#1,{{\jnfont J.\ Phys.\ G}{\bf #1},}
\def\CTP#1,{{\jnfont Commun.\ Theor.\ Phys.\ }{\bf #1},}
\def\ZPC#1,{{\jnfont Z.\ Phys.\ C }{\bf #1},}
\def\JHEP#1,{{\jnfont JHEP \ }{\bf #1},}
\def\Rv{\not{\hbox{\kern-1pt $R$}}}
\def\p{\not{\hbox{\kern-3pt $p$}}}

\begin{document}
\preprint{\parbox{1.2in}{\noindent arXiv:}}

\title{New physics effects on top quark spin correlation and
   polarization at the LHC: a comparative study in different models}

\author{Junjie Cao$^1$, Lei Wu$^2$ and Jin Min Yang$^2$
        \\~ \vspace*{-0.3cm}}
\affiliation{
$^1$ Physics Department, Henan Normal University, Xinxiang 453007, China \\
$^2$ Key Laboratory of Frontiers in Theoretical Physics,
     Institute of Theoretical Physics,
     Academia Sinica, Beijing 100190, China
     \vspace*{1.5cm}}

\begin{abstract}
Extensions of the Standard Model often predict new chiral
interactions for top quark, which will contribute to top quark spin
correlation and polarization in $t\bar{t}$ production at the LHC. In
this work, under the constraints from the current Tevatron
measurements, a comparative study of the spin correlation and
polarization is performed in three new physics models: the minimal
supersymmetric model without R-parity (RPV-MSSM), the third-generation enhanced
left-right model and the axigluon model. We find that the
polarization asymmetry may be enhanced to the accessible level in
all these models while the correction to the spin correlation may 
be detectable in the axigluon model and the RPV-MSSM with $\lambda''$ couplings.
\end{abstract}


\maketitle

\section{INTRODUCTION}

Among the known elementary particles, top quark is distinguished for
its excessively large mass and therefore often speculated to be
sensitive to new physics. So far the Tevatron experiments have
measured some of its properties such as the $t\bar{t}$ cross section
$\sigma(t\bar{t})$ and the differential cross section in each bin of
$t\bar{t}$ invariant mass $M_{t\bar{t}}$ \cite{top-review-exp}. While
most of the measurements agree well with the Standard Model (SM)
predictions, its forward-backward asymmetry shows moderate deviation
 \cite{top-afb-exp} which may be a harbinger for new physics
 \cite{top-afb-th1,top-afb-th2,top-afb-th3,top-afb-th4,top-afb-th5}.
Although the Tevatron collider is still running to collect more
data, the top quark measurement will be limited by its small
statistics. The Large Hadron Collider (LHC), however, will copiously
produce top quarks, which provides good opportunities to precisely
measure top quark properties and also to probe new
physics \cite{top-review}.

To explore new physics effects on top quark processes, the spin
information of top quark may be utilized because top quark decays
rapidly before forming any hadronic bound state and its spin
information is thus preserved. Explicitly speaking, for the $t\bar
t$ production at the LHC, top quark is not polarized at the leading
order of the SM since the production proceeds mainly through the QCD
interaction and the parity-violating electroweak contributions  to
the spin polarization is negligibly small  \cite{ew-tt}. But on the
other side, some extensions of the SM may predict new
parity-violating interactions of top quark which can enhance the
polarization sizably  \cite{np-polarization1,np-polarization2}.
Therefore, the polarization asymmetry can serve as a sensitive probe
to new physics. In association with the spin polarization, the spin
correlation of $t$ and $\bar t$ in the $t\bar t$ production, which
can be readily generated through the parity-conserving QCD
interaction  \cite{NLO-spin}, may also be altered significantly by
the new interactions and so can be used as an additional way to
probe new physics \cite{spin-np}.

In the popular minimal supersymmetric standard model (MSSM), the
SUSY effect on top quark polarization in the $t\bar t$ production
mainly arises from the radiative correction induced by the
axial-vector couplings of gluinos and thus the effect is
small \cite{susytt-loop}. In the general left-right models, since the
predicted new neutral gauge bosons are usually at TeV scale and
their mediated contributions to $t\bar{t}$ production do not
interfere with corresponding QCD amplitudes \cite{LR}, it is unlikely
to induce sizable polarization effects on the $t\bar{t}$ production.
In the extended color interaction models such as the topcolor model
 \cite{topcolor}, the breaking of the new color gauge symmetry will
give rise a massive octet coloron with strong coupling to top quark.
But due to its vector interaction nature with fermions, the coloron
mainly affects the $t\bar{t}$ production rate instead of the
polarization asymmetry  \cite{coloron}. Based on above observation,
we investigate the top quark polarization asymmetry and the spin
correlation between $t$ and $\bar{t}$ in the $t\bar{t}$ production
at the LHC in another three models: the R-parity violating minimal
supersymmetric standard model (RPV-MSSM), the third-generation
enhanced left-right model (LR model) and the axigluon model. These
models affect the $t\bar{t}$ production by exchanging at tree level
the color-singlet sleptons ($\tilde l^i_L$) and/or the color-triplet
squarks ($\tilde d^k_R$), the color-singlet vector boson $Z^{'}$ and
the color-octet vector boson $G^{'}$ respectively. As shown below,
although these effects on the $t\bar{t}$ production cross section
may be quite small, they may be sizable in the top quark
polarization and the spin correlation.

In our calculation, we take the SM parameters as  \cite{pdg}
\begin{eqnarray}
m_t=172.5{\rm ~GeV},~m_{Z}=91.19 {\rm
~GeV},~\sin^{2}\theta_W=0.2228. ~\alpha_s(m_t)=0.1095,~\alpha=1/128,
\end{eqnarray}
and use the CTEQ6L1 \cite{cteq} parton distribution function with the
renormalization scale $\mu_R$ and the factorization scale $\mu_F$
set to be $m_t$. The following three quantities are considered:
\begin{itemize}
\item[(i)] The ratio $P_t$ defined by \cite{np-polarization2}
\begin{eqnarray}
P_{t}=\frac{(\sigma_{+-}+\sigma_{++})-(\sigma_{--}+\sigma_{-+})}{\sigma_{+-}+\sigma_{++}+\sigma_{--}+\sigma_{-+}},
\end{eqnarray}
where  the polarization states for the $\bar{t}$ quark ($\pm$ denote
the helicity) are summed over.
\item[(ii)] The ratio $A_{LR}$ defined by \cite{np-polarization2}
\begin{eqnarray}
A_{LR}=\frac{\sigma_{+-}-\sigma_{-+}}{\sigma_{+-}+\sigma_{-+}} ,
\end{eqnarray}
where the difference between helicity-unlike cross sections of
$t\bar{t}$ pair is considered.
\item[(iii)] The correction to the spin correlation of $t$ and $\bar t$
defined by \cite{NLO-spin}
\begin{eqnarray}
\delta C=\frac{C_{tot}-C_{SM}}{C_{SM}}
\end{eqnarray}
where
\begin{eqnarray}
C=\frac{(\sigma_{++}+\sigma_{--})-(\sigma_{+-}+\sigma_{-+})}{\sigma_{++}+\sigma_{--}+\sigma_{+-}+\sigma_{-+}}.
\end{eqnarray}
\end{itemize}
In calculating these quantities, we have included the SM
contribution. We require new physics predictions of
$\sigma(t\bar{t})$ and $M_{t\bar{t}}$ at the Tevatron to coincide
with their measured values at $2\sigma$ level \cite{cross
section,mtt}.

This paper is organized as follows. In Sec. II, III and IV we
calculate the spin polarization and the spin correlation in the
RPV-MSSM, the third-generation enhanced left-right model and the
axigluon model respectively. We investigate the characteristics of
the quantities in a comparative way. In Sec. V we discuss the
observability of the quantities and then draw our conclusion.

\section{spin polarization and spin correlation in the RPV-MSSM}

The most general superpotential of the MSSM consistent with the SM gauge symmetry and
supersymmetry contains $R$-violating interactions, which are given
by~ \cite{rpv}
\begin{equation}\label{poten}
{\cal W}_{\not \! R}=\frac{1}{2}\lambda_{ijk}L_iL_jE_k^c
+\lambda'_{ijk} L_iQ_jD_k^c
+\frac{1}{2}\lambda''_{ijk}\epsilon^{\alpha\beta\gamma}U_{i\alpha}^cD_{j\beta}^cD_{k\gamma}^c
+\mu_iL_iH_2,
\end{equation}
where $i,j,k$ are generation indices, $c$ denotes charge
conjugation, $\alpha$, $\beta$ and $\gamma$ are the color indices
with $\epsilon^{\alpha\beta\gamma}$ being the total antisymmetric
tensor,  $H_{2}$ is the Higgs-doublet chiral superfield, and
$L_i(Q_i)$ and $E_i(U_i,D_i)$ are the left-handed lepton (quark)
doublet and right-handed lepton (quark) singlet chiral superfields
respectively. The dimensionless coefficients $\lambda_{ijk}$
(antisymmetric in $i$ and $j$) and $\lambda'_{ijk}$ in the
superpotential are $L$-violating couplings, while $\lambda''_{ijk}$
(antisymmetric in $j$ and $k$) are $B$-violating couplings.

\begin{figure}[htb]
\epsfig{file=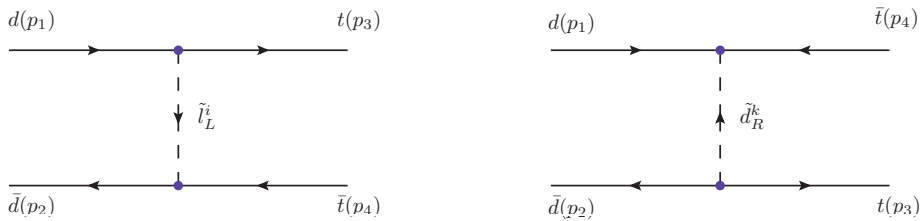,width=13cm} \vspace{-0.5cm} \caption{Feynman
diagrams contributing to $t\bar{t}$ production in the RPV-MSSM with
$\tilde l^i_L$ and $\tilde d^k_R$ denoting $i$-th generation
left-handed slepton and $k$-th generation right-handed squark
respectively.} \label{fig1}
\end{figure}

\begin{table}[t]
\caption{The upper bounds on the couplings $\lambda^{'}_{i31}$
($i=1,2,3$) and $\lambda^{''}_{31k}$ ($k=2,3$)  \cite{bounds-review}.
}
 \begin{tabular}{ccl} \hline
 couplings &~~~~~~~~~~~~~~bounds &~~~~~~~~~~~~~~~~~sources\\
 \hline
$\lambda^{\prime}_{131}$ &~~~~~~~~~~~~~ $0.03~m_{\tilde u^i_L}/(100$ GeV)&~~~~~~~~~~~~~~~~~$Q_W(Cs)$\\
$\lambda^{\prime}_{231}$ &~~~~~~~~~~~~~ $0.18~m_{\tilde{d}^k_L}/(100$ GeV)&~~~~~~~~~~~~~~~~~~$\nu_\mu q$ \\
$\lambda^{\prime}_{331}$ &~~~~~~~~~~~~~ $0.26~m_{\tilde{d}^k_R}/(100$ GeV)&~~~~~~~~~~~~~~~~~~$K\to\pi\nu\bar{\nu}$ \\
$\lambda^{''}_{31k}$     &~~~~~~~~~~~~~ $0.97~m_{\tilde d^k_R}/(100$ GeV)  &~~~~~~~~~~~~~~~~~~$R^{Z}_{l}$ \\
$\lambda^{''}_{31k}$     &~~~~~~$1.25$  &~~~~~~~~~~~~~~~~~~perturbativity\\
\hline
 \end{tabular}\label{dlambda}
 \end{table}


\begin{figure}[htb]
\epsfig{file=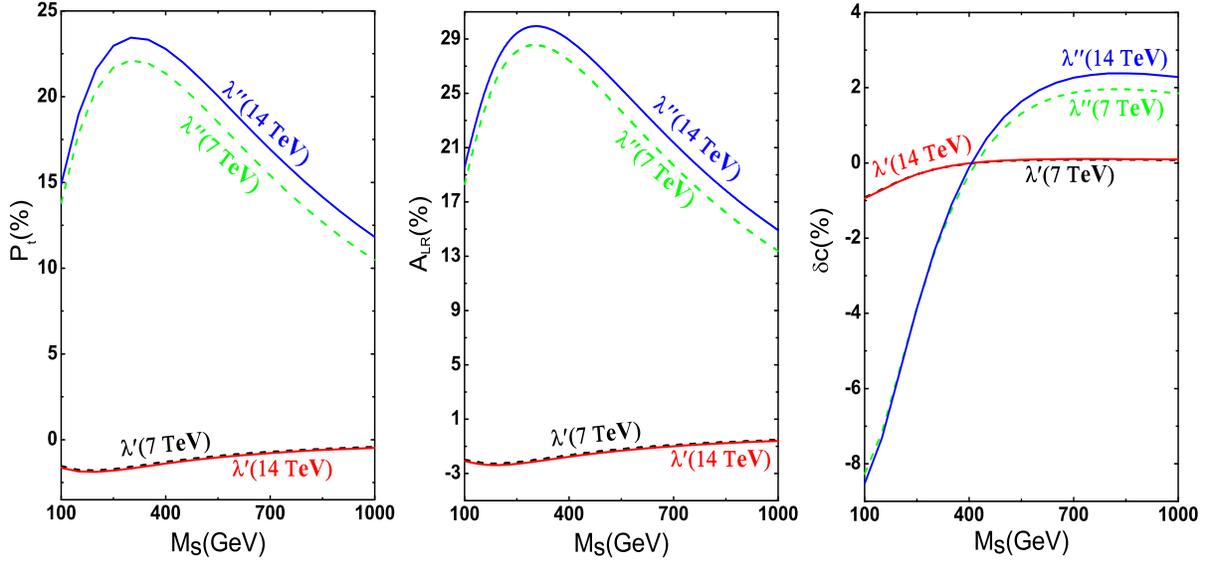,width=16cm,height=8cm} 
\vspace{-1.0cm}
 \caption{The contributions of $\lambda^\prime_{i31}$
(lower curves) and $\lambda^{''}_{31k}$ (upper curves) to $P_{t}$,
$A_{LR}$ and $\delta C$ at the LHC with $\sqrt{s}$=7 and 14 TeV,
which are represented by the solid and dash lines. } \label{fig2}
\end{figure}

\begin{figure}[htb]
\epsfig{file=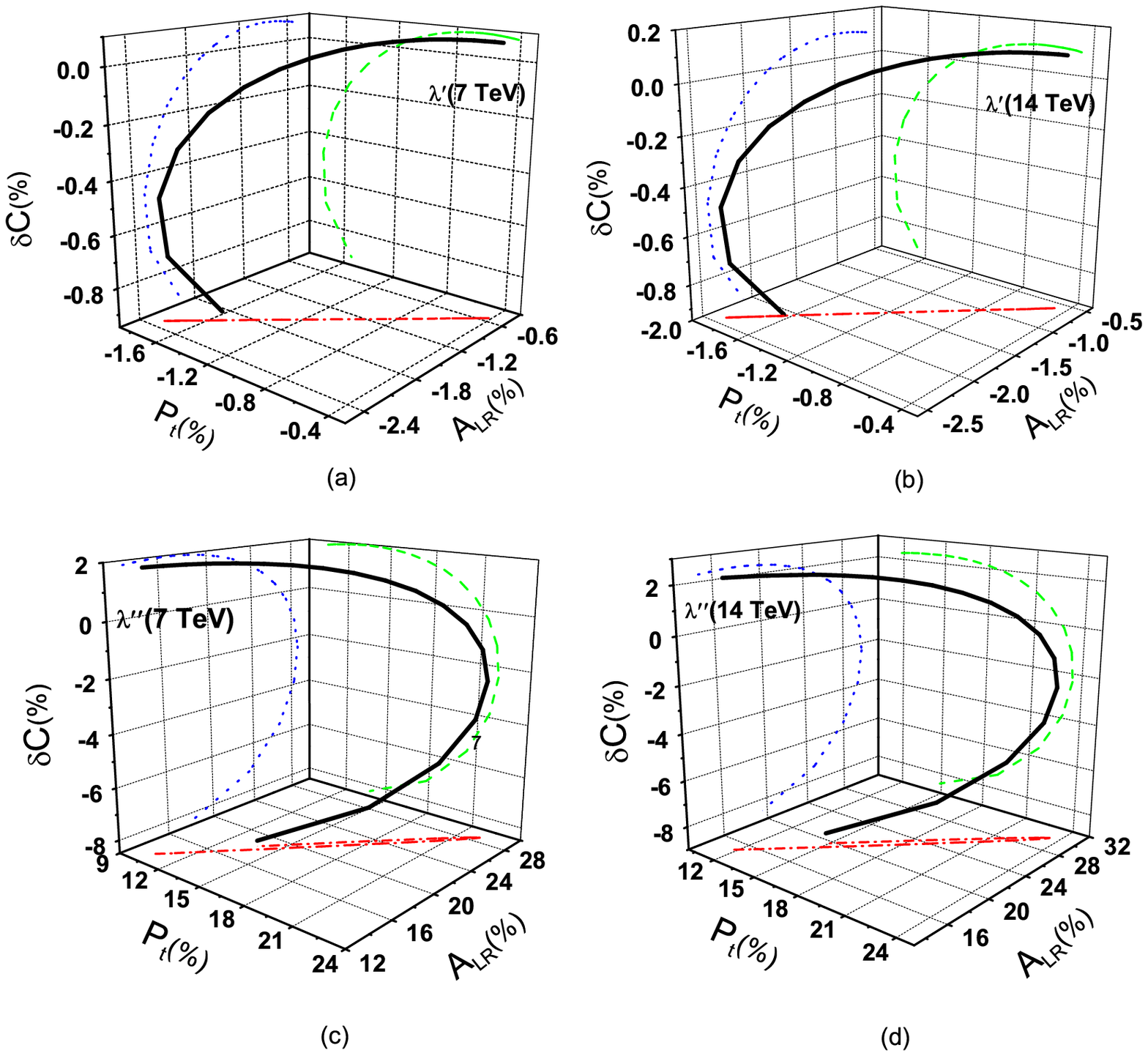,width=15cm,height=15cm} \vspace{-0.5cm}
 \caption{The correlation among $P_t$, $A_{LR}$ and $\delta C$ for
$\lambda^{'}$ and $\lambda^{''}$ respectively at the LHC with
$\sqrt{s}$=7 and 14 TeV. The projections(the blue, red and green
lines) on different planes are also shown.} \label{fig3}
\end{figure}

In Eq.(6) both $\lambda^\prime$ and $\lambda^{\prime \prime}$ terms
can induce new chiral interactions of top quark, which, in terms of
the four-component Dirac notation, are given by
\begin{eqnarray}
{\cal L} &=& \lambda^{\prime}_{ijk}
  \tilde l^i_L  \overline{d^k_R} u^j_L - \frac{1}{2}\lambda^{\prime\prime}_{ijk} [ \tilde d^{k*}_R \bar
u^{i}_R d^{jc}_L+\tilde d^{j*}_R\bar u^i_Rd^{kc}_L ]+h.c.
\end{eqnarray}
These interactions contribute to the $t\bar{t}$ production by the
diagrams shown in Fig.\ref{fig1} and their corresponding amplitudes
are
\begin{eqnarray}
M_{d\bar{d} \to t \bar{t}}^{RPV}|_{\lambda^\prime} &=&
-i\delta_{\alpha\rho}\delta_{\beta\sigma}|\lambda_{i31}^{'}|^{2}
\frac{\bar{u}(t)P_R
u(d)\bar{v}(d)P_Lv(t)}{(p_1-p_3)^2-m_{\tilde{l}_{iL}}^{2}} , \label{singlet} \\
M_{d\bar{d} \to t \bar{t}}^{RPV}|_{\lambda^{\prime \prime}} &=&
-i\varepsilon_{\beta\rho\lambda}\varepsilon_{\sigma\alpha\lambda}|\lambda_{31k}^{''}|^{2}
\frac{\bar{u}(t)\gamma_\mu P_R v(t)\bar{v}(d)\gamma^\mu
P_Ru(d)}{2[(p_1-p_4)^2-m_{\tilde{d}_{kR}}^{2}]} \label{triplet} ,
\end{eqnarray}
where $\alpha$, $\beta$, $\rho$, $\sigma$ and $\lambda$ are color
indices of the quarks and squarks and the sum over the generation
indices $i$ and $k$ is assumed. Since the amplitudes depend on the
coefficients $\lambda^\prime_{i31}$ and $\lambda^{''}_{31k}$, to
reasonably estimate the RPV-MSSM effect on the top pair production
we consider their upper bounds from different measurements
 \cite{rp2,bounds-review}, which are summarized in Table I. This
table shows that the bounds are proportional to squark mass and for
squark as heavy as $1$ TeV, the coefficients may be of
${\cal{O}}(1)$. In our discussion, we assume all the
squarks/sleptons degenerate in mass ($m_{\tilde l^i_L}=m_{\tilde
d^k_R}=M_{s}$) and take the largest values of $\lambda^\prime_{i31}$
and $\lambda^{''}_{31k}$ for given $M_s$ to maximize the RPV-MSSM
effects. We require $M_s$ larger than about $100$ GeV, which
corresponds to the mass bound from the LEP search for the
sparticles \cite{LEP}.

Our results indicate that although the amplitudes interfere with the
QCD amplitude $d\bar{d} \to g^\ast \to t \bar{t}$, their effects on
the $t\bar{t}$ production cross section is only few percent
 \cite{rp1} and thus unobservable at the LHC; but on the other hand,
since the interactions violate parity, their effect on spin
polarization may be sizable. The latter conclusion becomes obvious
from Fig.\ref{fig2} where we plot the three quantities $P_t$,
$A_{LR}$, and $\delta C$ versus $M_s$ for the LHC with $\sqrt{s}=7,
14$ TeV respectively.  This figure shows that, due to the largeness
of $\lambda^{\prime \prime}$, the values of $P_t$, $A_{LR}$ and
$\delta C$ can reach $23\%$, $30\%$ and $-8.5\%$ respectively for
$\sqrt{s}=14$ TeV. In addition, we note that even if one further
requires the RPV-MSSM to explain the top quark forward-backward
asymmetry measured at the Tevatron at $2 \sigma$ level, which favors
the squark mass from $250$ GeV to $400$ GeV \cite{top-afb-th3}, the
polarization asymmetry $P_t$ and $A_{LR}$ may still be tens percent
and thus be observable at the LHC (see Table II in Sec. V).

Since in the RPV-MSSM, the three quantities depend on the same couplings,
i.e. $\lambda^{'}$ or $\lambda^{''}$, their values should be correlated,
which may be utilized to distinguish different models.
In Fig.3 we show such correlation for $\lambda^{'}$ and $\lambda^{''}$, respectively.
We see that the curves for $\lambda^{'}$ is very
different from those for $\lambda^{''}$.

About the RPV-MSSM, two points should be noted. Although the
coupling $\lambda^{\prime\prime}_{31i}$ in the RPV model can be
severely constrained by the $n-\bar{n}$ oscillation, the upper bound
are dependent on the squark mass and other SUSY parameters, such as
the chargino mass $m_{\tilde{w}}$ and the soft SUSY breaking
parameters$A_q$ In the Ref.[3], they obtain the constrains for the
scenario $m_{\tilde{q}}=m_{\tilde{w}}=A$, which is not relevant for
our calculation. The other is we use the mass bounds of the
sparticles from the LEP instead of from the Tevatron. The reason is
we are considering R-parity violating case while the Tevatron
results are valid only for the $R$-conserving case.

\section{ spin polarization and spin correlation in the LR model}

So far various left-right symmetric models have been proposed to
understand the origin of parity violation and neutrino masses. These
models generally predict new gauge bosons which contribute to
$t\bar{t}$ production at tree level  \cite{LR}. However, due to the
heaviness of these bosons, they are hard to induce any significant
effect on the $t\bar{t}$ production when the constraints from
Tevatron are considered. Here we focus on a special left-right
symmetric model called the third-generation enhanced left-right
model, which still allows relatively light new gauge bosons and
neutral flavor changing interaction at tree level \cite{hexg}. This
model is based on the gauge group $SU(3)_C \times SU(2)_L\times
SU(2)_R \times U(1)_{B-L}$ with gauge couplings $g_3$, $g_L$, $g_R$
and $g$ respectively, and the key feature of this model is  the
right-handed gauge bosons corresponding to $SU(2)_R$ group couple
only to the third-generation fermions. After the mixings of the
right handed bosons with the SM bosons and the light quarks in the
SM with top quark, the $t \bar{t}$ production will be affected by
the modified SM gauge interactions and also by additional
contributions from new gauge bosons. In our analysis, we only
consider the potentially large contribution from neutral gauge
interactions, which are given by \small
\begin{eqnarray}
{\cal L}_Z &=& -{g_L\over 2 \cos\theta_W} \bar q \gamma^\mu (g_V -
g_A \gamma_5) q (\cos\xi_Z Z_\mu - \sin\xi_Z Z^\prime_\mu)
\nonumber\\
&+& {g_Y\over 2} \tan\theta_R ({1\over 3} \bar q_L \gamma^\mu q_L+
{4\over 3} \bar u_{Ri} \gamma^\mu u_{Ri} -{2\over 3} \bar
d_{Ri}\gamma^\mu d_{Ri})
(\sin\xi_Z Z_\mu + \cos\xi_Z Z^\prime_\mu)\nonumber\\
&-& {g_Y\over 2} (\tan\theta_R + \cot\theta_R) ( \bar u_{Ri}
\gamma^\mu V^{u*}_{Rti} V^{u}_{Rtj}u_{Rj} - \bar d_{Ri} \gamma^\mu
V^{d*}_{Rbi} V^{d}_{Rbj} d_{Rj}) (\sin\xi_Z Z_\mu + \cos\xi_Z
Z^\prime_\mu) \label{neucoup}
\end{eqnarray}
\normalsize where $\tan \theta_R = g/g_R$, $g_Y = g \cos\theta_R =
g_R \sin\theta_R$, $\xi_{Z}$ is the mixing angle between $Z_R$ and
$Z_0$, $q$ and $q_L$ denote any quarks, $V^{u,d}_{Rij}$ are the
unitary matrices which rotate the right-handed quarks $u_{Ri}$ and
$d_{Ri}$ from interaction basis to mass eigenstates and the repeated
generation indices $i$ and $j$ are summed.

\begin{figure}[tbp]
\epsfig{file=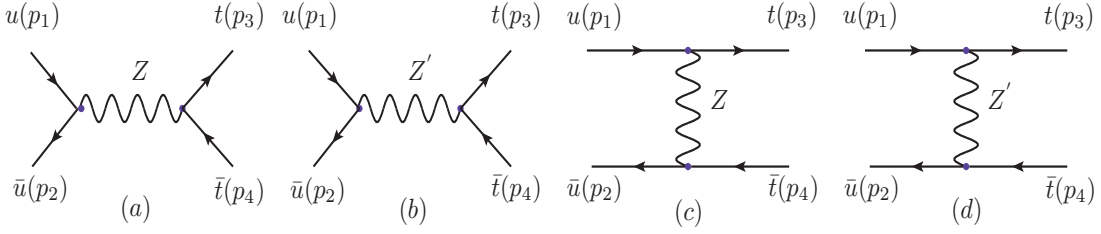,width=15cm, height=3.3cm} \vspace*{-0.7cm}
\caption{Feynman diagrams contributing to $t\bar{t}$ production in
the third-generation enhanced left-right model.} \label{fig4}
\end{figure}

Eq.(\ref{neucoup}) indicates that the $Z^\prime \bar{u}_i u_j$
interaction is strong if $g_R \gg g_Y$ or equally $\cot \theta_R \gg
1$. In  \cite{hexg} nearly diagonal mixing matrices $V^d_{R}$ and
$V^u_{R}$ were introduced to avoid large flavor-changing neutral
currents. In practice, this requirement may be relaxed since sizable
$u_R-t_R$ mixing with the other flavor mixings suppressed can still
satisfy the constraints \cite{top-afb-th2}. Here we emphasize that
this pattern of flavor mixing does not necessarily mean the up-top
element in up-type quark mass matrix $M_u$ is much larger than other
off-diagonal elements. For example, assuming $(V_{R}^u)_{ut} = 0.2$,
$(V_{R}^u)_{ct} =0 $ and $(V_{R}^u)_{uc}=0$, we numerically solve
the equation $V^{u \dagger}_R M_u^\dagger M_u V^u_R = M^2_{diag}$
with $M^2_{diag}=diag\{m_u^2,m_c^2,m_t^2\}$, and we find it possible
that $(M_{u})_{ct}$ is several times larger than $(M_{u})_{ut}$.
With the non-vanishing $u_R-t_R$ mixing, the top pair production may
proceed by the $t$-channel diagrams shown in Fig.\ref{fig4} (c-d),
and unlike the s-channel contribution in Fig.\ref{fig4} (a-b), this
t-channel contribution will interfere with the QCD process $ u
\bar{u} \to g^\ast \to t \bar{t}$, so is more important than the
s-channel contribution.

\begin{figure}[tbp]
\epsfig{file=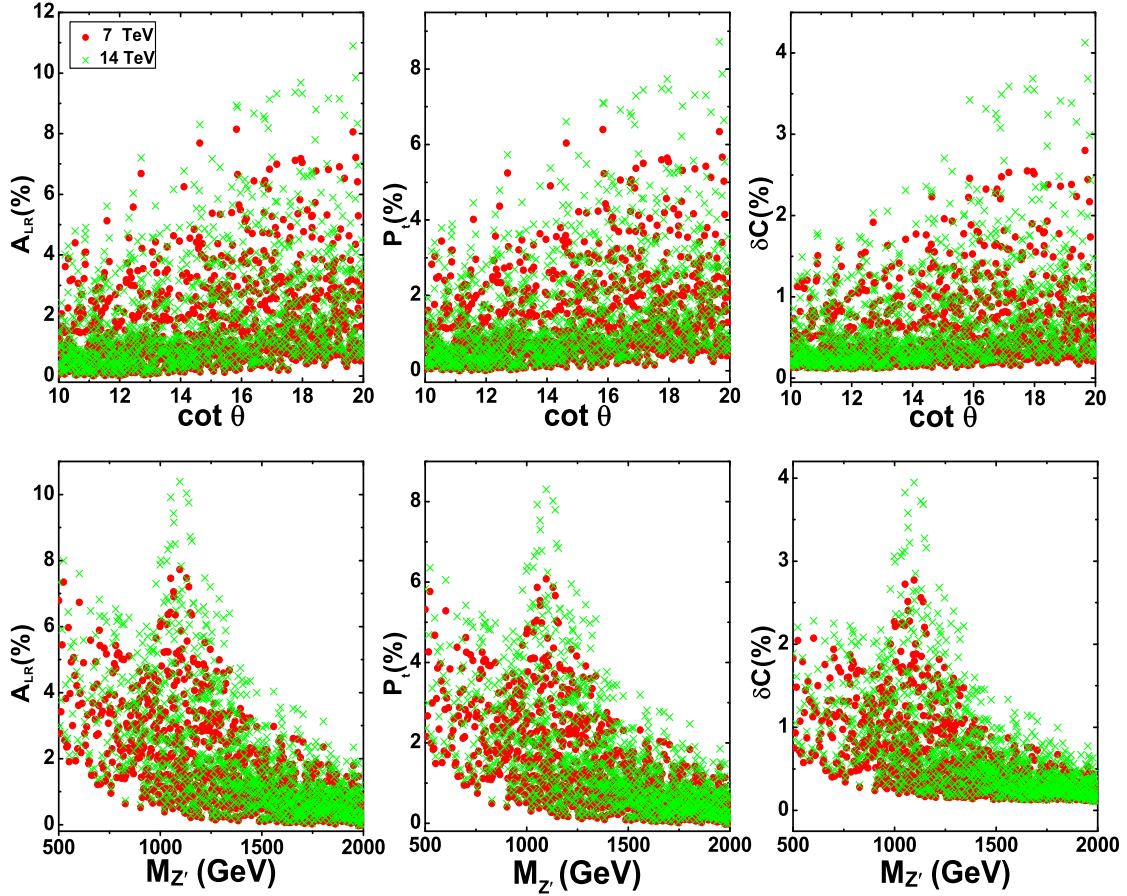,width=15cm, height=12cm} 
\vspace*{-0.5cm}
\caption{Scatter plots of the quantities $P_t$, $A_{LR}$ and $\delta
C$ as a function of $M_{Z^{'}}$ or $\cot{\theta}_{R}$
in the LR model. Constraints from the Tevatron
measurements on $\sigma (t \bar{t})$ and $M_{t\bar{t}}$ are imposed.
The bullets (red) and crosses (green) denote  respectively 
the center-of-mass energy $\sqrt{S}=7$ TeV and 14 TeV at LHC. } 
\label{fig5}
\end{figure}

\begin{figure}[htb]
\epsfig{file=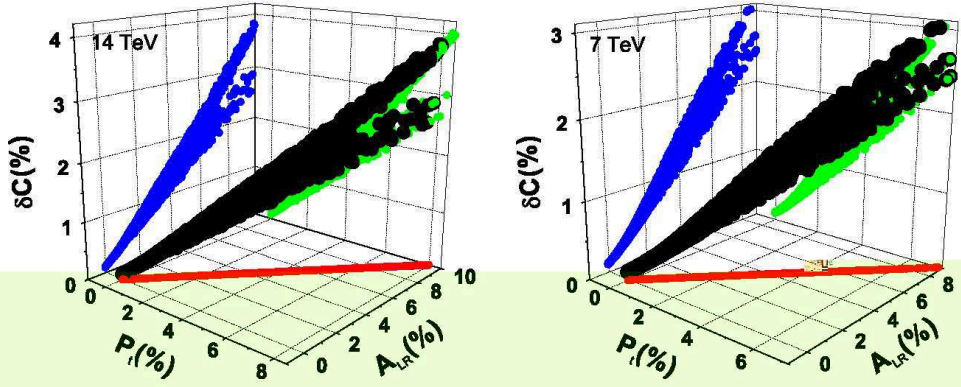,width=13cm} 
\vspace{-0.5cm}
\caption{Same as Fig.5, but showing the correlation of different
polarization observables at the LHC with $\sqrt{s}$=7 and 14 TeV.
The projections (the blue, red and green dots) on different planes
are also shown.} \label{fig6}
\end{figure}

The amplitudes corresponding to Fig.\ref{fig4} are
given by
\small
\begin{eqnarray}
M_a &=& i\delta_{\alpha\beta}\delta_{\rho\sigma}\left(\frac{e}{2c_w
s_w}\right)^{2}
\frac{\bar{u}(t)\gamma_{\mu}[g_{ZL}^{t}P_L+g_{ZR}^{t}P_R]
v(t)\bar{v}(u)\gamma^{\mu}[g_{ZL}^{u}P_L+g_{ZR}^{u}P_R]u(u)}{(p_1+p_2)^2-m_{Z}^{2}},  \label{amp1} \\
M_b &=& i\delta_{\alpha\beta}\delta_{\rho\sigma}\left(\frac{e}{2c_w
s_w}\right)^{2}
\frac{\bar{u}(t)\gamma_{\mu}[g_{Z^{'}L}^{t}P_L+g_{Z^{'}R}^{t}P_R]
v(t)\bar{v}(u)\gamma^{\mu}[g_{Z^{'}L}^{u}P_L+g_{Z^{'}R}^{u}P_R]u(u)}{(p_1+p_2)^2-m_{Z^{'}}^{2}-i\Gamma_{Z^{'}}m_{Z^{'}}}, \\
M_c &=& i\delta_{\alpha\rho}\delta_{\beta\sigma}\left(\frac{e}{2c_w
s_w}\right)^{2}\left[\xi_Z s_w
(\cot{\theta_R}+\tan{\theta_R})\right]^{2}|V_{Rtu}^{u}V_{Rtt}^{u}|^{2}
\frac{\bar{u}(t)\gamma_{\mu}P_R u(u)\bar{v}(u)\gamma^{\mu}P_R v(t)}{(p_1-p_3)^2-m_{Z}^{2}}, \\
M_d &=& i\delta_{\alpha\rho}\delta_{\beta\sigma}\left(\frac{e}{2c_w
s_w}\right)^{2}\left[s_w
(\cot{\theta_R}+\tan{\theta_R})\right]^{2}|V_{Rtu}^{u}V_{Rtt}^{u}|^{2}
\frac{\bar{u}(t)\gamma_{\mu}P_R u(u)\bar{v}(u)\gamma^{\mu}P_R
v(t)}{(p_1-p_3)^2-m_{Z^{'}}^{2}}, \label{amp4}
\end{eqnarray}
\normalsize where $\Gamma_{Z^\prime}$ is the $Z^\prime$ width
obtained by considering all decay channels of $Z^\prime$,
$s_w=\sin{\theta_{w}}$, $c_w=\cos{\theta_{w}}$, and the coupling
coefficients $g_{ZL}^{t,u}$, $g_{ZR}^{t,u}$, $g_{Z^{'}L}^{t,u}$ and
$ g_{Z^{'}R}^{t,u}$ are defined as
\begin{eqnarray}
g_{ZL}^{t,u} &=& 1- \frac{4}{3} s_{w}^{2} - \frac{1}{3} s_w \tan{\theta_R} \xi_Z, \\
g_{ZR}^{u} &=& -\frac{4}{3} s_{w}^{2} - \frac{4}{3} s_w \tan{\theta_R} \xi_Z, \\
g_{ZR}^{t} &=& -\frac{4}{3} s_{w}^{2} - \frac{1}{3} s_w
\tan{\theta_R} \xi_Z+ s_w \cot{\theta_R} \xi_Z, \label{coup1}\\
g_{Z^{'}L}^{t,u} &=& (1 - \frac{4}{3} s_{w}^{2} ) \xi_Z + \frac{1}{3} s_w \tan{\theta_R},  \label{coup2} \\
g_{Z^{'}R}^{u} &=& -\frac{4}{3} s_{w}^{2} \xi_Z + \frac{4}{3} s_w \tan{\theta_R}, \\
g_{Z^{'}R}^{t} &=& - \frac{4}{3} s_{w}^{2} \xi_Z + \frac{1}{3} s_w
\tan{\theta_R} \xi_Z - s_w \cot{\theta_R} .
\end{eqnarray}

The amplitudes in Eq.(\ref{amp1}-\ref{amp4}) involve the parameters
$\xi_z$, $\cot{\theta}_{R}$, $M_{Z^{'}}$ and $(V^u_R)_{ut}$. About
the parameters $\xi_{Z}$ and $m_{Z^{\prime}}$, the oblique $T$
parameter and perturbative requirement will give constraints on
$\xi_{Z}$ (versus $m_{Z^{\prime}}$)  \cite{hexg}. However, such
constraints are obtained under the requirement to explain the
$b$-quark forward-backward asymmetry $A^{b}_{FB}$, which, of course,
can be relaxed if we give up the explanation of $A^{b}_{FB}$.
Furthermore, for the $t\bar{t}$ productions, the main contributions
are independent of the parameter $\xi_{Z}$  \cite{top-afb-th4}.
Therefore, the constraints from the $T$ parameter are almost
irrelevant to our numerical study. We note that the constraints from
CDF search for $Z^\prime$  \cite{zprime} and from the global fitting
of the electroweak precision data  \cite{ewzprime} are invalid here
since these constraints arise mostly from the processes involving
the fermions of the first two generations. So far the most pertinent
bound comes from $e^{+}e^{-} \to b\bar{b}$ at LEP-II, which requires
$M_{Z^\prime} \gtrsim 460$ GeV for $\cot \theta_R \ge 10$
 \cite{hexg}. In our numerical calculation, we scan over following
parameter regions:
\begin{eqnarray}
500 {\rm ~GeV} \le M_{Z^{\prime}} \le 2000 {\rm ~GeV}, \quad 0 \le
\xi_Z \le 0.02, \quad  10 \le \cot{\theta}_{R} \le 20, \quad 0.1 \le
(V^{u}_R)_{ut} \le 0.2 \nonumber
\end{eqnarray}

Requiring the LR predictions for $\sigma(t\bar{t})$ and
$M_{t\bar{t}}$ at Tevatron to coincide with their measured values at
$2\sigma$ level, we display in Fig.5 the surviving samples on the
planes of the polarization asymmetries versus $\cot \theta_R$ or
$M_{Z^\prime}$ respectively. This figure shows that large $P_t$,
$A_{LR}$ and $\delta C$ come from the region where $\cot \theta_R$
is large. This can be understood by that in the LR model, the
dominant contribution comes from the diagram (d) of Fig.4, which is
proportional to $ ( \cot \theta_R + \tan \theta_R )^2$. This figure
also shows that the samples with $500<M_{Z^{'}}<900$ GeV affect
little on the quantities. We checked that these samples are strongly
constrained by the Tevatron measurements of $\sigma(t\bar{t})$ and
$M_{t\bar{t}}$.

In Fig.6 we display the correlation of $P_t$, $A_{LR}$ and $\delta
C$ for the surviving samples. We see that these three quantities are
proportional to each other, which is quite different from the
RPV-MSSM case.

Finally, we note that for most of the surviving samples,  the top
quark forward-backward asymmetry $A_{FB}$ measured at the Tevatron
lies within $2\sigma$ region around its experimental central
value \cite{top-afb-th3}. So even if we take the asymmetry as a
constraint, our results for the spin polarization asymmetry still
hold.

\section{spin polarization and spin correlation in the axigluon model}
\begin{figure}[htb]
\epsfig{file=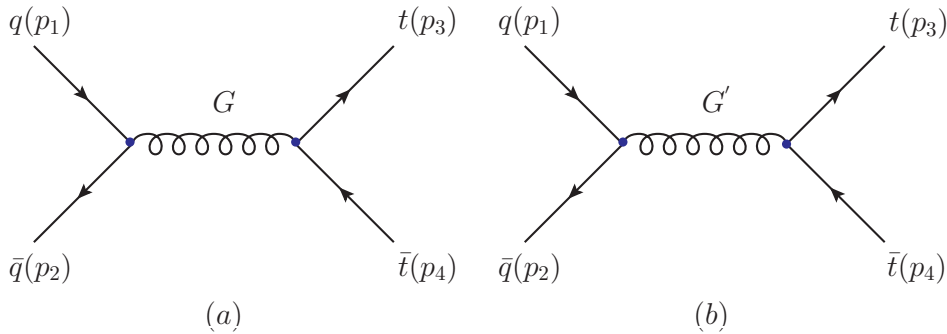,width=13cm} \vspace{-0.5cm} \caption{Feynman
diagrams contributing to $t\bar{t}$ production in the axigluon model
with $G$ and $G^{'}$ denoting gluon and axigluon respectively.}
\label{fig7}
\end{figure}

The third model we are considering is the chiral color model
 \cite{axigluon1} which enlarges the color group $SU(3)_C$ to
$SU(3)_L \bigotimes SU(3)_R$ with gauge couplings $g_L$ and $g_R$
respectively. The spontaneous breaking of this enlarged symmetry to
$SU(3)_C$ will produce massive color octet called axigluon.
Depending on charge assignments of quarks under the $SU(3)_L
\bigotimes SU(3)_R$  group, there are many variants of the chiral
color models. In this paper, we focus on a special one which was
utilized to explain the $2\sigma$ deviation of the top quark
forward-backward asymmetry \cite{top-afb-th4,top-afb-th5}. The key
feature of this model is the quarks in the third and the first two
generations are assigned with different chirality in the  $SU(3)_L
\bigotimes SU(3)_R$ group, and consequently, the couplings of the
axigluon with quarks are given by
\begin{eqnarray}
g^{q}_{V}=g^{t}_{V}=-g_{s}\cot{2\theta},~~
g^{q}_{A}=-g^{t}_{A}=-g_{s}\csc{2\theta},   \label{gvga}
\end{eqnarray}
where $\theta$ is the rotation angle relating gluon $G$ and axigluon
$G^\prime$ to the interaction eigenstates $G_1$ and $G_2$ by
\begin{eqnarray}
G=\cos\theta G_1+ \sin\theta G_2, \quad \quad G^{'} =\sin\theta
G_1-\cos\theta G_2.
\end{eqnarray}
The value of $\theta$ is determined by the gauge couplings, $\cot
\theta = \tan^{-1}(g_L/g_R)$, and the perturbativity and the
condition for fermion condensation require it vary from $14^{0}$ to
$45^{0}$.

\begin{figure}[htb]
\epsfig{file=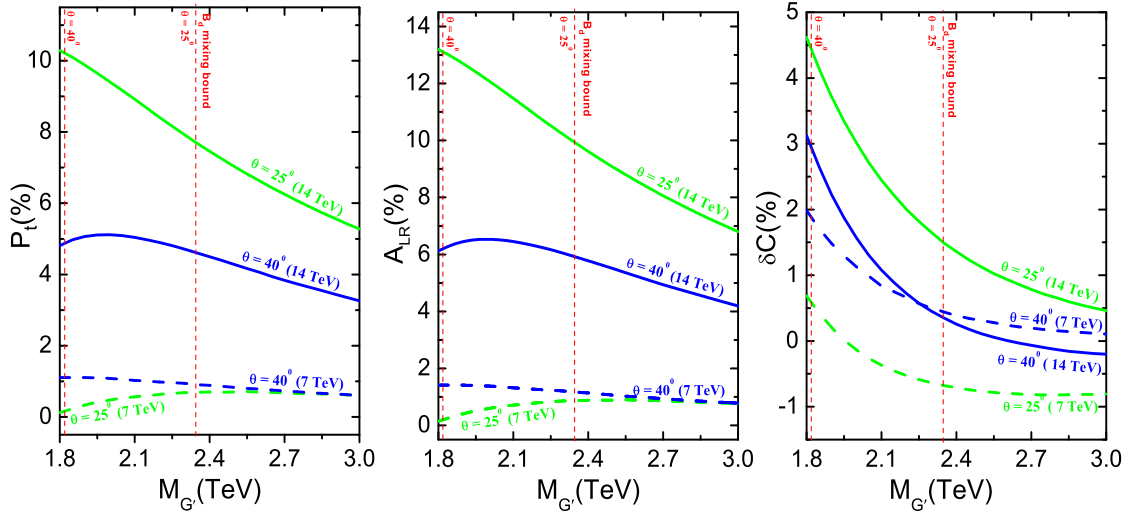,width=15cm,height=7cm} 
\vspace*{-0.5cm}
\caption{The
contribution of axigluon to top spin polarization and correlation at
the LHC with $\sqrt{s}$=7, 14 TeV. The vertical dashed lines indicate
the lower bounds on axigluon mass from $B_{d}-\bar{B_{d}}$ mixing  \cite{top-afb-th5}.} 
\label{fig8}
\end{figure}

In the axigluon model, the Feynman diagrams contributing to
$t\bar{t}$ production are displayed in Fig.7 and the corresponding
amplitudes are
\begin{eqnarray}
M_a &=& iT^{a}_{\beta\alpha}T^{a}_{\rho\sigma}g^{2}_{s}
\frac{\bar{u}(t)\gamma_{\mu}v(t)\bar{v}(q)\gamma^{\mu}u(q)}{(p_1+p_2)^2}, \\
M_b &=& iT^{a}_{\beta\alpha}T^{a}_{\rho\sigma}
\frac{\bar{u}(t)\gamma_{\mu}[g_{V}^{t}+g_{A}^{t}\gamma_5]
v(t)\bar{v}(q)\gamma^{\mu}[g_{V}^{q}+g_{A}^{q}\gamma_5]u(q)}{(p_1+p_2)^2
-m_{G^{'}}^{2}-i\Gamma_{G^{'}}m_{G^{'}}}  
\label{axigluon-amp}
\end{eqnarray}
where $g_s$ is the usual $SU(3)_C$ strong coupling, and
$\Gamma_{G^\prime}$ is the $G^\prime$ width obtained by assuming the
axigluon decays only to the SM quarks  \cite{top-afb-th2,top-afb-th4}.
With the Tevatron constraints $\sigma(t \bar{t})$ and
$M_{t\bar{t}}$, we show the dependence of $P_t$, $A_{LR}$ and
$\delta C$ on $M_{G^\prime}$ in Fig.\ref{fig8}. This figure
indicates that large predictions of these quantities come from the
case of light $G^{'}$ and small $\theta$, and for $\theta=25^{0}$,
$m_{G^\prime}=1.8 {\rm TeV}$ and $\sqrt{s}=14$ TeV, $P_t$, $A_{LR}$
and $\delta C$ can reach $10.3 \%$, $13.2 \%$ and $4.6\%$
respectively. 

Fig.\ref{fig8} shows that the
magnitude of the effects is quite sensitive to the axigluon mass.
About the axigluon mass, recently the direct search at 
the LHC gave a lower bound $m_{G'}> 1.52{\rm TeV}$  \cite{cms}, 
which, however, is for the flavor universal case and not applicable 
to our flavor non universal model. In Fig.\ref{fig8} we display 
an indirect bound $m_{G^{\prime}}\sin(2\theta)>1.8$ TeV
from the $B_{d}-\bar{B_{d}}$ mixing  \cite{top-afb-th5}.
However, such a bound was derived under an assumption that the 
flavor mixing between the left-handed down-type quarks is approximated 
as the SM CKM matrix  \cite{mass constrains}.
We see from Fig.\ref{fig8} that if we adopt such an assumption
to impose this bound, then the axigluon effects will be rather limited.

\begin{figure}[htb]
\epsfig{file=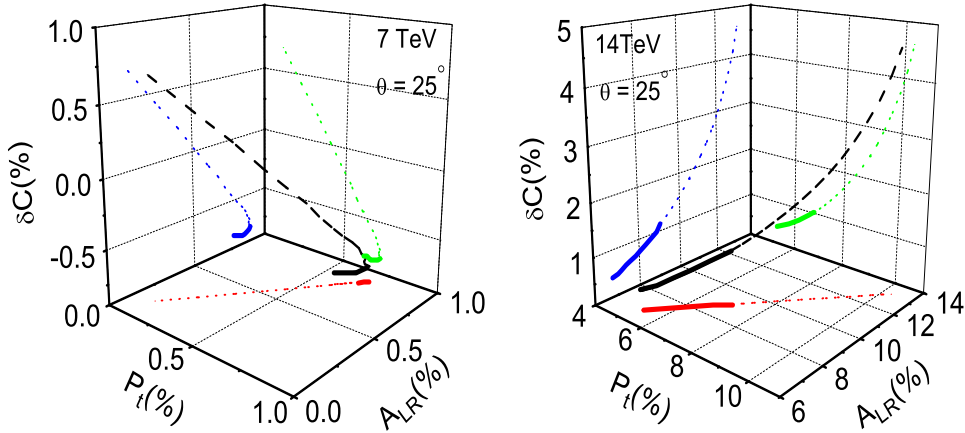,width=13cm,height=6cm}
\vspace{-0.5cm}\caption{The correlation among $P_t$, $A_{LR}$ and
$\delta C$ in the axigluon model for the LHC with $\sqrt{s}$=7,14
TeV. Projections on different planes are also shown.
For each curve, the solid part satisfies the lower bound on the 
axigluon mass from $B_{d}-\bar{B_{d}}$ mixing  \cite{top-afb-th5}.} 
\label{fig9}
\end{figure}

In Fig.9 we fix $\theta=25^{0}$ and display the correlation of
$P_t$, $A_{LR}$ and $\delta C$ at the LHC with $\sqrt{s}=7, 14$ TeV
respectively. Comparing with the results in the RPV-MSSM with $\lambda^{''}$,  
we see that the three quantities have different correlation behavior.
\begin{figure}[htb]
\epsfig{file=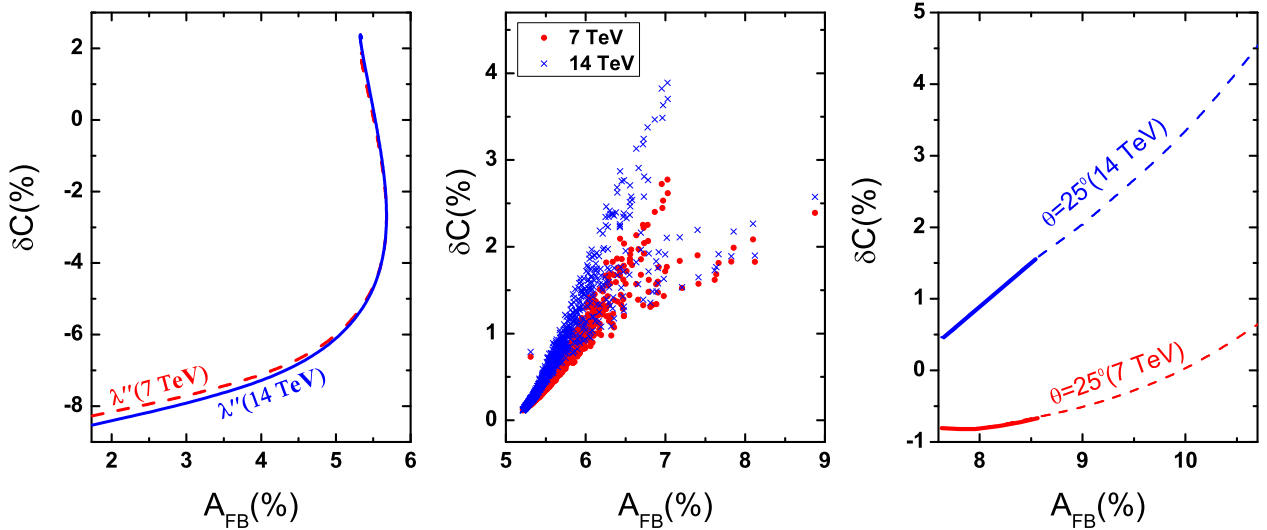,width=17cm}
\vspace*{-1.5cm}
\caption{The
correlation between $A_{LR}$ and $\delta C$ in the $\lambda^{''}$ ,
LR and axigluon models for the LHC with $\sqrt{s}$=7,14 TeV.
For the axigluon results displayed in the right frame, 
the solid part of each curve satisfies the lower bound on the 
axigluon mass from $B_{d}-\bar{B_{d}}$ mixing  \cite{top-afb-th5}.}
\label{fig10}
\end{figure}

Finally, in Fig.10 we display the correlation between the spin
correlation and the top quark forward-backward asymmetry in the
$\lambda^{''}$ RPV-MSSM, the LR and axigluon models, which can 
alleviate the $A^{t}_{FB}$ deviation. It can be seen that the correlation 
behaviors are quite different for different models.

\section{Discussion and Conclusion}

In order to discuss the observability of the asymmetries $P_t$ and
$A_{LR}$, we calculate the statistical significance $N_S$ defined in
 \cite{np-polarization2} and list their values in Table II for the
optimum case of $P_t$ and $A_{LR}$ in the three models with an
integrated luminosity $\cal{L}$ =1 $fb^{-1}$. From the results of
Fig.\ref{fig2},\ref{fig5},\ref{fig8} and Table II, we have following
observations:
\begin{itemize}
\item[(1)] The LHC with $\sqrt{s}=14 {\rm TeV}$ usually predicts 
larger $N_S$ than that with $\sqrt{s}=7 {\rm TeV}$. For the RPV-MSSM, 
the $\lambda^{'}$-induced asymmetries are obviously unobservable at 
the LHC, while the $\lambda^{''}$-induced asymmetries may be detectable.
For the specific left-right model and the axigluon model, $P_t$ and
$A_{LR}$ may also be large enough to be detected at the LHC. On the
other hand, if the effects are not observed in the future, the stronger 
bounds can be imposed on the models.

\item[(2)] Among the three models, the RPV MSSM and the axigluon model 
can allow for $\delta C$ larger than $4\%$ in magnitude, which
may be observable at the LHC  \cite{spin-atlas}.
\end{itemize}

In summary, in this work we considered new physics effects on the
top quark spin polarization and correlation at the LHC in three
models: the RPV-MSSM, the specific left-right model and the axigluon
model. We find that, due to the introduction of new parity-violating
interactions of top quark, the polarization asymmetries $P_t$ and
$A_{LR}$ can reach the observable level in all these models, 
while for the spin correlation 
the RPV MSSM and the axigluon model can cause observable effects 
at the LHC.

\section*{Acknowledgement}
We thank Tao Han for discussions.
This work was supported in part by HASTIT under grant No.
2009HASTIT004, by the National Natural Science Foundation of China
(NNSFC) under grant Nos. 10821504, 10725526, 10635030, 10775039,
11075045 and by the Project of Knowledge Innovation Program (PKIP)
of Chinese Academy of Sciences under grant No. KJCX2.YW.W10.

\begin{table}
\caption{\small The maximal statistical significance $N_S$ (defined
in  \cite{np-polarization2}) for $P_{t}$ and $A_{LR}$ at the LHC with
an integrated luminosity of 1 fb$^{-1}$.}
\begin{tabular}{|l|l|l|l|l|l|l|l|l|} \hline
~  & \multicolumn{2}{c|}{RPV-MSSM ($\lambda^{'}$)} &
\multicolumn{2}{c|}{RPV-MSSM ($\lambda^{''}$)}
 & \multicolumn{2}{c|}{LR Model ($Z^{'}$)} & \multicolumn{2}{c|}{Axigluon Model ($g^{'}$)} \\ \cline{2-9}
 ~  &~~$P_t$ &~~$A_{LR}$~ &~~$P_t$ &~~$A_{LR}$~ &~~$P_t$ &~~$A_{LR}$~ &~~$P_t$ &~~$A_{LR}$~\\
\hline ~~7 TeV &~~1.7 $\sigma$ &~~1.9 $\sigma$ &~~29.1 $\sigma$
&~~36.5 $\sigma$ &~~8.8 $\sigma$ &~~9.9 $\sigma$ &~~1.71 $\sigma$
&~~1.95 $\sigma$
\\ \hline ~14 TeV &~~3.1 $\sigma$  &
~~3.5 $\sigma$  &~~59.2 $\sigma$  &~~68.3 $\sigma$ &~~18.4 $\sigma$ &~~20.6 $\sigma$ &~~26.5 $\sigma$ &~~32.8 $\sigma$ \\
\hline
\end{tabular}
\end{table}

\end{document}